\documentclass[prb,twocolumn,showpacs,superscriptaddress]{revtex4-1}

\usepackage{amsmath}
\usepackage{amssymb}

\usepackage{longtable}%
\usepackage{dcolumn}
\usepackage{bm}
\usepackage{color}

\usepackage{graphicx}
\usepackage{dcolumn}
\usepackage{bm}
\usepackage{amsmath}
\usepackage{tabu}
\usepackage{multirow}
\usepackage[latin1]{inputenc}
\usepackage{subfigure}
\usepackage{braket}
\usepackage{soul}
\setstcolor{red}
\usepackage{xcolor}
\usepackage{verbatim}
\usepackage[dvipdfm, pdfstartview=FitH, CJKbookmarks=true, bookmarksnumbered=true,
bookmarksopen=true, colorlinks, pdfborder=001, linkcolor=blue, anchorcolor=blue, citecolor=blue]{hyperref}

\begin{document}
\preprint{Regular article}


\title{Synthesis, phase stability, structural and physical properties of 11-type iron chalcogenides}

\author{Sahana R\"o{\ss}ler}
\email{roessler@cpfs.mpg.de}
\affiliation{Max Planck Institute for Chemical Physics of Solids,
N\"othnitzer Stra\ss e 40, 01187 Dresden, Germany}
\author{Cevriye~Koz}
\affiliation{Max Planck Institute for Chemical Physics of Solids,
N\"othnitzer Stra\ss e 40, 01187 Dresden, Germany}
%
%
%
%
%
\author{Steffen Wirth}
\affiliation{Max Planck Institute for Chemical Physics of Solids,
N\"othnitzer Stra\ss e 40, 01187 Dresden, Germany}
\author{Ulrich Schwarz}
\affiliation{Max Planck Institute for Chemical Physics of Solids,
N\"othnitzer Stra\ss e 40, 01187 Dresden, Germany}

\date{\today}
\begin{abstract}

%
%
%
  This article reviews recent experimental investigations on two binary Fe-chalcogenides: FeSe and Fe$_{1+y}$Te. The main focus is on synthesis, single crystal growth, chemical composition, as well as on the effect of excess iron on structural, magnetic, and transport properties of these materials. The structurally simplest Fe-based superconductor Fe$_{1+x}$Se with a critical temperature $T_c \approx$ 8.5~K undergoes a tetragonal to orthorhombic phase transition at a temperature $T_s \approx$ 87~K. No long-range magnetic order is observed down to the lowest measured temperature in Fe$_{1+x}$Se. On the other hand, isostructural Fe$_{1+y}$Te displays a complex interplay of magnetic and structural phase transitions in dependence on the tuning parameter such as excess amount of Fe or pressure, but it becomes a superconductor only when Te is substituted by a sufficient amount of Se. We summarize experimental evidence for different competing interactions and discuss related open questions. 

 \end{abstract}

%
%

\maketitle   

\section{Introduction}
The discovery of superconductivity in LaFeAsO$_{1-x}$F$_{x}$ with a $T_{c}$ = 26 K by Kamihara $et~al.$ \cite{kam2008} unveiled a new field of research generally referred to \textit{Fe-based~superconductors} (Fe-SC). Several new superconducting phases have been discovered with highest $T_{c}$ of 56~K achieved
so far in bulk samples. \cite{Ren2008,Wan2008}. Among the different families of Fe-SC, FeSe (11-type chalcogenide) has the
simplest crystal structure \cite{Hsu2008}. The atomic structure belongs to the tetragonal $P4/nmm$ space
group and consists of edge-sharing FeSe$_{4}$ tetrahedra, which form layers orthogonal to the $c$-axis.
The bulk $T_{c}$ of Fe$_{1+x}$Se is 8.5~K at ambient conditions, with the superconducting properties being
extremely sensitive to the amount of excess Fe \cite{Mc2009}. The $T_{c}$ of FeSe can be increased to as high as 37 K by the application of hydrostatic pressure \cite{Miz2008,Med2009,Mar2009,Ima2009,Sid2009}. This
makes Fe$_{1+x}$Se a member of the high-$T_{c}$ class of compounds. Fe$_{1+x}$Se does not order magnetically, but spin fluctuations were detected by nuclear magnetic resonance (NMR) measurements \cite{Ima2009}.
The
spin fluctuations are found to be strongly increased near $T_{c}$, and applied pressure
seems to enhance the spin fluctuations along with the superconducting transition temperature \cite{Ima2009}. On the other hand,
$T_{c}$ of FeSe can also be increased by Te substitution of Se, up to a maximum of $T_{c} \approx$ 15 K
for Fe$_{1+y}$Se$_{0.5}$Te$_{0.5}$ \cite{Yeh2008,Fan2008,Ros2010,Che2015a}. The superconducting volume fraction decreases with increasing Te and no superconductivity has been found so far in bulk samples of the end-member
Fe$_{1+y}$Te.

It is also important to mention that a monolayer of FeSe on a SrTiO$_{3}$ substrate becomes superconducting with $T_{c}$ in the range of 65-100~K \cite{Yan2012,He2013,Tan2013,Ge2015,Fan2015}. 
Angle resolved photoemission spectroscopic (ARPES) measurements observed its Fermi surface being distinct from other iron-based superconductors, consisting only of electron-like pockets near the zone corner without indication of any hole-like Fermi surface around the $\Gamma$ point \cite{He2013,Liu2013}. The high-$T_{c}$ in the FeSe monolayer likely originates from an electron doping by the substrate at the interface thereby modifying the Fermi surface \cite{Tan2013,Bang2013}. Although the properties of FeSe monolayers are extremely interesting, here we limit ourselves to bulk materials only. Further, this article gives by no means a complete review of the vast existing literature on Fe-chalcogenides, it only provides an overview with emphasis on the composition and homogeneity of the samples. 
\section{F\lowercase{e}S\lowercase{e}}
In the literature, there has been some confusion concerning the nomenclature of the tetragonal FeSe. In this article, following the early publications of the binary phase diagram of FeSe \cite{Sch1979,Oka1991}, the tetragonal FeSe is referred to as $\beta$-FeSe.   

\subsection{Synthesis and characterization}

    Although the binary phase diagram of Fe$_{1+x}$Se was known for about three decades \cite{Sch1979,Oka1991}, the interest in this compound rose tremendously when superconductivity was found in FeSe$_{1-\delta}$ by Hsu \textit{et~al}. \cite{Hsu2008}. These polycrystalline samples were synthesized by mixing high-purity Se and Fe powder and carrying out the reaction at 700$^\circ$C, which is slightly above the boiling point of Se (for the detailed synthesis procedure see Ref. \cite{Hsu2008}). Superconductivity was reported for highly Se deficient samples FeSe$_{0.82}$. The x-ray diffraction pattern of these samples displayed several impurity phases as well as unknown phases. This work was followed by a combined x-ray and neutron diffraction study which reported a composition for the superconductor as FeSe$_{0.92}$ \cite{Mar2008}. Both these compositions fall outside of the composition reported in  Ref. \cite{Sch1979,Oka1991} for the binary phase diagram of $\beta$-FeSe. 
		
		Alternatively, McQueen \textit{et~al}. \cite{Mc2009} considered stoichiometric FeSe samples. They used two different temperature steps of 750$^\circ$C and 1025$^\circ$C, respectively, before cooling the sample to 420$^\circ$C, followed by subsequent quenching of the sample to -13$^\circ$C \cite{Mc2009,Wil2009}. This procedure yielded FeSe samples free of impurities. Further, McQueen \textit{et~al}. \cite{Mc2009} reported that Fe$_{1.01}$Se and Fe$_{1.02}$Se are superconductors with $T_{c}$ = 8.5~K and 5 K, respectively, whereas Fe$_{1.03}$Se is not a superconductor at all. A more detailed comparison \cite{Pom2009} on samples synthesized by the above mentioned two different routes concluded that a stable phase exhibiting superconductivity at 8~K exists in a
narrow range of selenium concentration FeSe$_{0.974\pm0.005}$ ($\approx$ Fe$_{1.027\pm0.005}$Se). More recently, Koz \textit{et~al.} \cite{Koz2014} synthesized FeSe by following the procedure described in Ref. \cite{Mc2009} and starting from the nominal compositions of the Fe:Se ratio in the range 0.98 - 1.02. The lattice parameters obtained from the x-ray diffraction of these samples are presented in Fig.~1. It can be seen in Fig.1 that the stability range of Fe$_{1+x}$Se  
narrows the nominal composition range to $0.00 \leq x \leq 0.01$. Thus, the composition of iron selenide lies in the range Fe$_{1.00}$Se to Fe$_{1.01}$Se, and is referred to hereafter as $\beta$-FeSe.

\begin{figure}[t]%
\includegraphics*[width=8 cm, clip]{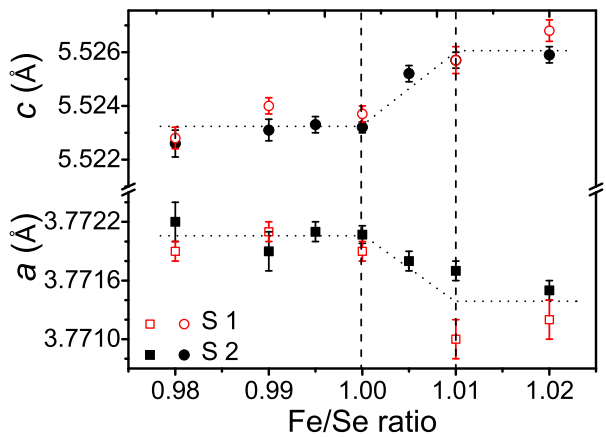}
\caption{%
  Lattice parameters of  $\beta$-FeSe at room temperature for the
nominal compositions Fe:Se between 0.98 and 1.02. Labels S1 and S2 represent two different series of synthesis under the same conditions \cite{Koz2015}.}
\label{Fig1}
\end{figure}

\subsection{Single crystal growth}
Different methods have been attempted for the growth of single crystals of FeSe. These methods include vapor self transport \cite{Pat2009}; alkali-halide-flux growth with KCl (or KBr) \cite{Mok2009,Gor2012}, NaCl/KCl \cite{Zha2009}, LiCl/CsCl \cite{Hu2011}, or KCl/AlCl$_{3}$ \cite{Cha2013,Boe2013}; the Bridgman technique \cite{Yan2011}; the traveling-solvent float zone technique \cite{Ma2014}; chemical vapor transport with I$_{2}$ \cite{Kar2012}, TeCl$_{4}$ \cite{Har2010}, or FeBr$_{3}$ \cite{Wu2013}; a cubic-anvil high-pressure technique \cite{Wu2013}; and chemical vapor transport with AlCl$_{3}$ \cite{Koz2014}. Many of these techniques did not yield high-quality single crystals of FeSe. For instance, for single crystals growth with only alkali-halide flux \cite{Zha2009,Hu2011} as well as by the Bridgman technique \cite{Yan2011}, impurities like $\delta$-Fe$_{1-y}$Se and $\alpha$-Fe were reported. In addition, FeSe crystals grown by the Bridgman method \cite{Yan2011} showed a dendrite-like morphology. Further, the magnetization $M(H)$ hysteresis loops suggested the presence of ferromagnetic impurities \cite{Pat2009,Hu2011}. 

\begin{figure}[t]%
\includegraphics*[width=6 cm, clip]{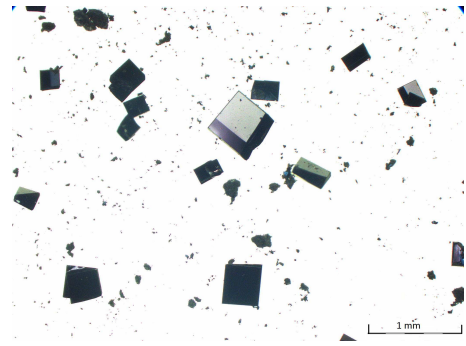}
\caption{%
  Single crystals of FeSe grown by chemical vapor transport.}
\end{figure}
\begin{figure}[b]%
\includegraphics*[width=8 cm, clip]{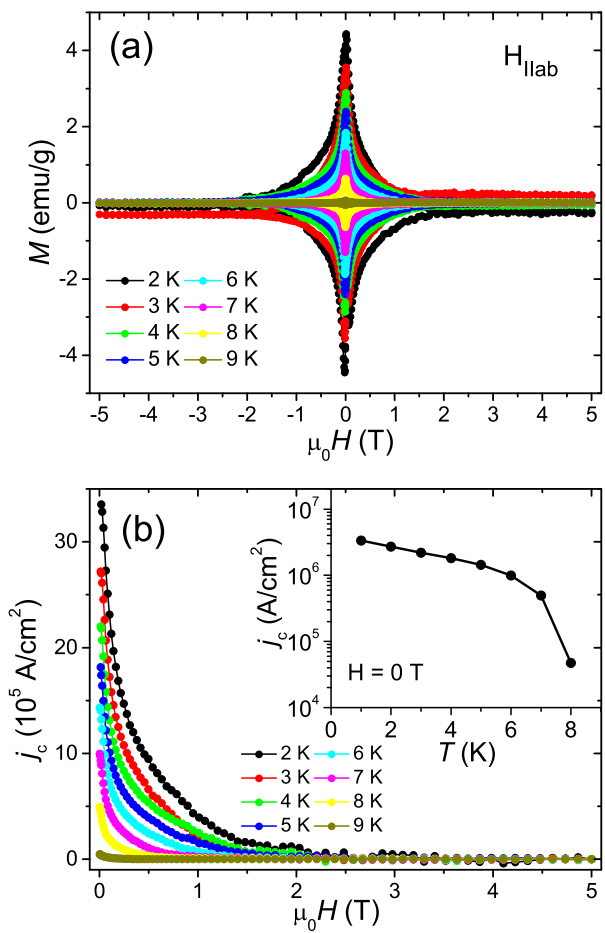}
\caption{%
  (a) Magnetization isotherms of a $\beta$-FeSe single crystal below 10 K as
a function of magnetic field. Field is applied parallel to the $ab$-plane. (b) Critical current
density vs. magnetic field for $\beta$-FeSe with $\mu_{0}H$ parallel to the $ab$-plane. The inset shows the
temperature dependence of the critical current density at $\mu_{0}H$ = 0 T on logarithmic scale,
obtained from (b) \cite{Koz2015}.}
\label{Fig2}
\end{figure}

  Among the above-mentioned different methods, two procedures turned out to be successful in growing high-quality single crystals of FeSe, either by utilizing a eutectic mixture of KCl and AlCl$_{3}$ \cite{Cha2013,Boe2013,Lin2011,Huy2014} or chemical vapor transport performed with AlCl$_{3}$ \cite{Koz2014}. In the former method, B\"ohmer \textit{et~al.} \cite{Boe2013} took powders of Fe and Se in an atomic ratio of 1.1:1. The powders were put in an evacuated SiO$_{2}$ ampule together with an eutectic mixture of KCl and AlCl$_{3}$. The ampoule was heated to 390$^\circ$C 
on one end while the other end was kept at 240$^\circ$C. After 28.5 days isometric FeSe crystals with tetragonal morphology were extracted at the colder end of the ampule. The level of impurities in these single crystals was found to be less than 500 ppm. The single crystals grown by this method were utilized in several sophisticated physical experiments \cite{Mal2014,Ter2014,Shi2014,Kas2014,Zha2015,Kn2015,Wat2015,Boe2015,Wat2015b,Wata2015} to explore the electronic properties of FeSe.  

 Alternatively, Koz \textit{et~al.} \cite{Koz2014} grew single crystals of FeSe by chemical vapor transport using only very small quantities of AlCl$_{3}$ as transport reagent. In their method, 1~g of stoichiometric FeSe powder was taken along with 20~mg of AlCl$_{3}$ in an evacuated quartz ampule. The ampule was placed in a two-zone furnace at temperatures $T_{1}$ = 400$^\circ$C and $T_{2}$ = 300$^\circ$C. The growth was typically carried out for 2-3 months. Finally, the ampule was quenched in water to room temperature. The product contained plate-like crystals with tetragonal morphology (see Fig. 2) with edge lengths up to 0.5~mm and a maximum thickness of 0.01 mm. The crystals were repeatedly washed with ethanol to remove any remaining condensed gas phase, dried under vacuum, and stored in argon-filled glove boxes. By extending the growth time to 1 year, larger single crystals with dimensions up to 4 $\times$ 2 $\times$ 0.03 mm$^{3}$ were grown \cite{Koz2014}. The single crystals were characterized by x-ray diffraction, wave length dispersive x-ray spectroscopy, and electron diffraction. The results of these experiments proved a high quality of the single crystals and hence, the amount of impurities present in the sample was assumed to be below the detection limit of these physical measurements. Further, topography measured by scanning tunneling microscopy displayed large surfaces free of impurities and defects \cite{Ros2015}. In addition to these methods, magnetization measurements at room temperature were utilized to estimate the amount of elemental Fe in the samples. According to these estimates, the best polycrystalline samples contained between 100
and 300 ppm elemental iron \cite{Koz2014}. The bulk nature of superconductivity in the crystals was confirmed by specific heat measurements \cite{Koz2014}. Typical values for the residual resistivity ratio $\rho_{300~\mathrm{K}}$/$\rho_{12~\mathrm{K}}$ are found to be in the range 22-27 for these crystals.  

The single crystals grown by the above method are free of magnetic impurities as can be seen from the magnetization isotherms $M(H)$.
Exemplary $M(H)$ curves measured for a $\beta$-FeSe single crystal below 10 K 
are plotted in Fig. 3(a) \cite{Koz2015}. A typical superconducting
magnetic hysteresis curve is observed below $T_{c}$ = 8.5~K. There is no indication of a second peak
(fishtail effect) within the measured field range. In contrast, such a fishtail feature was observed for
tetragonal FeTe$_{0.6}$Se$_{0.4}$ \cite{Yad2009}. From the $M(H)$ loops, the critical current density, $j_{c}$,
can be calculated by using the Bean critical state model \cite{Bea1962,Bea1964}:
\begin{equation*} 
j_{c} = 20 \frac{\Delta M}{a(1-a/3b)}
\end{equation*}
in which $\Delta M = M\!\!\!\downarrow - ~ M\!\!\!\uparrow$. Here, $M\!\!\!\downarrow$ and $M\!\!\!\uparrow$ are the $M(H)$ measured with
increasing and decreasing field, respectively, whereas $a$ and $b$ ($b > a$) are the dimensions
of the rectangular cross section of the crystal normal to the applied field. Here, the
field is applied perpendicular to the $c$-axis. For the present crystal, $a$ = 0.003~cm and $b$ = 0.13~
cm. Since the measured crystal is a very thin plate and the applied
field is parallel to the long axis, the demagnetization factor was assumed to be negligible. Fig. 3(b) presents the
critical current density at several temperatures as a function of field. The
calculated $j_{c}$ from our $M(H)$ curves at zero field is 10$^{6}$ A/cm$^{2}$ at 2 K, which is higher
than the previously reported values for $\beta$-FeSe \cite{Son2011,Lei2011,Oza2012,Ding2012} but similar to other iron based
superconductors \cite{Yad2009,Ding2012,Yan2008,Tae2009}. The inset of Fig. 3 (b) shows the critical
current density as a function of temperature. $j_{c}$ is uniformly decreasing up to 7~K but
still amounts to 10$^{4}$ A/cm$^{2}$ at 8~K.

\subsection{Structural phase transition and nematicity}

One of the features of the parent compounds of Fe-SC is that they undergo a tetragonal to orthorhombic structural phase transition at a temperature $T_s$. The structural phase transition can either occur simultaneously with a spin density wave transition at a N\'eel temperature $T_{N}$ = $T_s$, or they are split such that $T_{N} < T_s$, see for details Refs. \cite{Lum2010,Joh2010,Ste2011}.  At $T_{s}$, the materials develop a large in-plane anisotropy in the resistivity \cite{Chu2010,Fis2011}, $i.e$., a spontaneous breaking of the $C_4$ rotational symmetry, while preserving the translational symmetry of the underlying electronic system.  This large in-plane electronic anisotropy is termed nematicity and the terms ``structural transition'' and ``nematic transition'' have become synonyms in the literature. However, the origin of the nematic state is highly controversial \cite{Fer2014} because such a transition can be induced by either phonons, or orbital or spin fluctuations. Understanding the exact nature of this phase transition is considered highly important for obtaining insight into the mechanism of superconductivity as well as the symmetry of the superconducting order parameter \cite{Fer2014}. 

\begin{figure}[t]%
\includegraphics*[width=8.5 cm, clip]{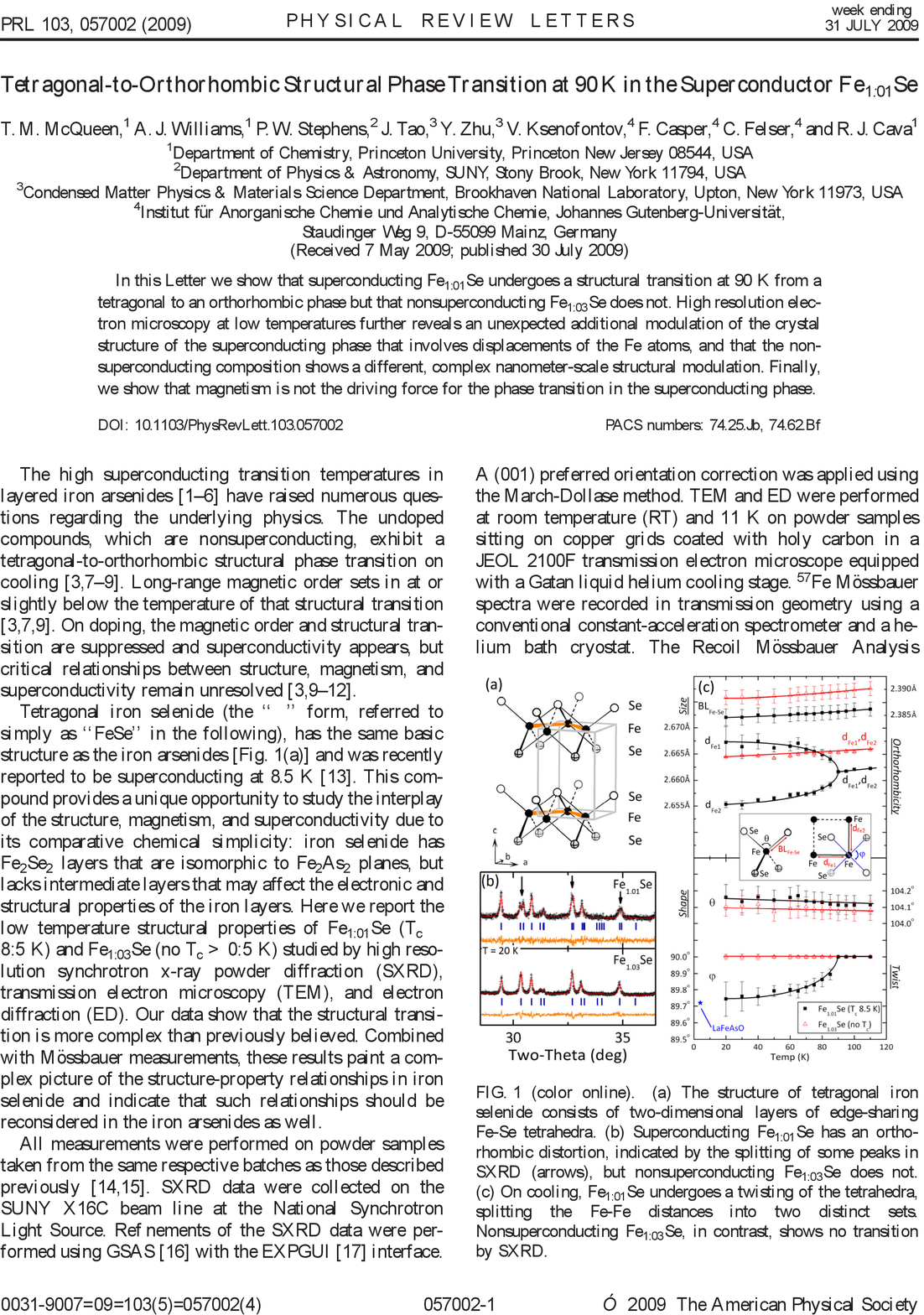}
\caption{%
  (a) The crystal structure of tetragonal FeSe consists of two-dimensional layers of edge-sharing
Fe-Se tetrahedra. (b) Superconducting Fe$_{1.01}$Se has an orthorhombic
distortion, indicated by the splitting of some peaks in
SXRD (arrows), but nonsuperconducting Fe$_{1.03}$Se does not.
(c) On cooling, Fe$_{1.01}$Se undergoes a twisting of the tetrahedra,
splitting the Fe-Fe distances into two distinct sets.
Nonsuperconducting Fe$_{1.03}$Se, in contrast, shows no transition
by SXRD. Reproduced from Ref. \cite{Mc2009b} after obtaining permission from the authors, copyright (2009) by American Physical Society.}
\end{figure}    

$\beta$-FeSe is deemed to be an ideal candidate material for addressing the origin of the nematic transition in Fe-SC.
At room temperature, it has a tetragonal structure with a space group $P4/nmm$. It exhibits a transition to an orthorhombic $Cmma$ phase at $T_s$= 87~K \cite{Mar2008,Pom2009,Mc2009,Mil2009}, but does not order magnetically down to 2~K. The structural parameters for Fe$_{1.01}$Se and Fe$_{1.03}$Se obtained by McQueen \textit{et~al.} are reproduced in Fig.~4 from Ref. \cite{Mc2009b}. In Fig 3(a), the tetragonal structure of FeSe is presented. The superconducting Fe$_{1.01}$Se has an orthorhombic
distortion, indicated by the splitting of some peaks in the synchrotron x-ray diffraction (SXRD) pattern, Fig. 4(b). The orthorhombic distortion is due to coherent twisting of the
upper and lower Se pairs away from the ideal angle of 90$^\circ$. It can be described by five parameters: the torsional
angle between the Se pairs ($\varphi$), two Fe-Fe distances ($d_\mathrm{Fe1}$
and $d_\mathrm{Fe2}$), the Fe-Se bond length (BL$_{\mathrm{Fe-Se}}$), and the upper
Se-Fe-Se angle ($\theta$) \cite{Mc2009b}. The temperature-dependence of these
parameters is shown in Fig. 4(c). It is interesting to note that the structural transition is found only in the superconducting Fe$_{1.01}$Se, but not in the non-superconducting Fe$_{1.03}$Se \cite{Mc2009b}. Therefore, an observation of this structural transition can be considered as a good test for the composition of FeSe samples. The single crystals of FeSe grown by KCl/AlCl$_{3}$ flux \cite{Boe2013} and chemical vapor transport with AlCl$_{3}$ \cite{Koz2014}, both displayed the orthorhombic phase transition.

The origin of the structural phase transition in FeSe nonetheless remains controversial. Although $\beta$-FeSe does not display a long-range magnetic order, signatures of spin-fluctuations have been detected in NMR \cite{Ima2009,Boe2015,Bae2015}, magnetic susceptibility \cite{Ros2015}, and inelastic neutron scattering experiments \cite{Rah2015,Wan2016}. The latter measurements, when performed at an energy of 13 meV did not show any variation in spin-fluctuations across the structural phase transition \cite{Rah2015}. The highly dispersive paramagnetic fluctuations found along ($\pi$,0) in the Fe-square lattice extending beyond 80 meV in energy suggested that FeSe is close to an instability towards ($\pi$,0) antiferromagnetism that is characteristic of the parent phases of the Fe-pnictide superconductors \cite{Rah2015}. In contrast, an inelastic neutron scattering experiment \cite{Wan2016} conducted at low energies of 2.5 meV on single crystalline FeSe found an onset of the dynamical spin correlation function $S(\bf{Q},\omega)$ at the orthorhombic phase transition. The temperature evolution of $S(\bf{Q},\omega)$ was compared with the orthorhombic distortion, $\delta(T)=(a-b)/(a+b)$, where $a$ and $b$ are the lattice parameters of FeSe. The results indicated that the enhancement of $S(\bf{Q},\omega)$ is coupled to the orthorhombic phase, which is consistent with the theoretical proposals of a nematic order driven by spin fluctuations \cite{Yu2015,Gla2015,Wang2015,Chu2015}.  

Alternatively, an orbital-driven nematic scenario was favored by the results based on NMR experiments \cite{Boe2015,Bae2015}, in which, unlike the neutron scattering, probe only momentum-integrated spin fluctuations at very low energies. The spin-lattice relaxation rate measured in those NMR experiments starts to diverge at temperatures significantly lower than the structural transition temperature of 87~K. Based on this fact, B\"ohmer \textit{et~al.} argued that the spin-fluctuations are not the driving force of the structural phase transition \cite{Boe2015}. Further, Baek \textit{et~al.} observed an order-parameter-like temperature dependence of the Knight shift when measured with magnetic field applied parallel to the crystallographic $a$ axis \cite{Bae2015}. Since the Knight shift is proportional to the orbital-order parameter, they concluded that the nematicity in FeSe is driven by orbital order. Several ARPES experiments reported a 50 meV splitting of $d_{xz}$ and $d_{yz}$ bands \cite{Shi2014,Zha2015,Wat2015,Nak2014,Suz2015}, which is considered as an evidence of nematicity. On the other hand, the first ARPES measurement on FeSe \cite{Mal2014} contradicted the above results by claiming that all observed ARPES spectral features can be explained by regular band structure calculations, therefore they do not provide any evidence for nematic order. Further, it was shown that even though a band splitting of 50 meV was found in the energy dispersive curves, by taking into account the intrinsic widths of these peaks, the possible remaining splitting was estimated to be in the order of 5~meV \cite{Bor2015}. Note that this is a reasonable estimation of an energy scale of the structural phase transition occurring at 87~K.

\begin{figure}
\includegraphics[width=8 cm,clip]{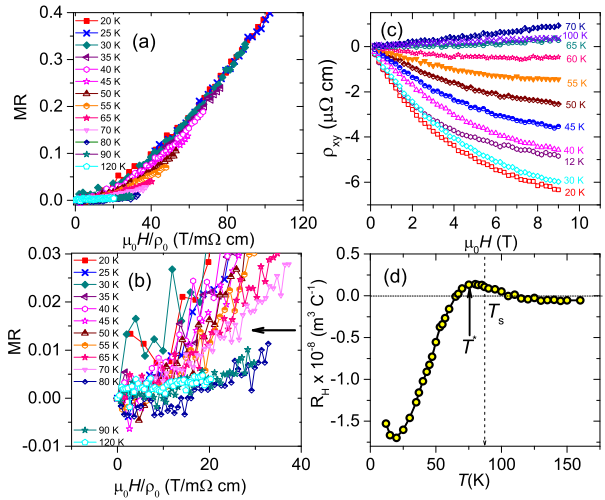}
\caption{(a) Kohler plot in the temperature range 20-120 K displaying the validity of Kohler's scaling of magnetoresistance (MR) below 30~K and above 70~K. (b) Same data as shown in panel (a), but on an enlarged scale. The horizontal arrow is placed to emphasize the deviation of MR from Kohler's scaling. (c) Hall resistivity $\rho_{xy}$ in the temperature range 12-100~K. (d) The temperature dependence of the initial Hall coefficient $R_{H\rightarrow0}$ displays a clear deviation from the compensated metal regime below $T^{*}\approx$ 75~K where spin fluctuations become important. Reproduced from Ref. \cite{Ros2015}, copyright (2015) by American Physical Society.} 
\end{figure}
\begin{figure}[t]
\includegraphics[width=8.5 cm,clip]{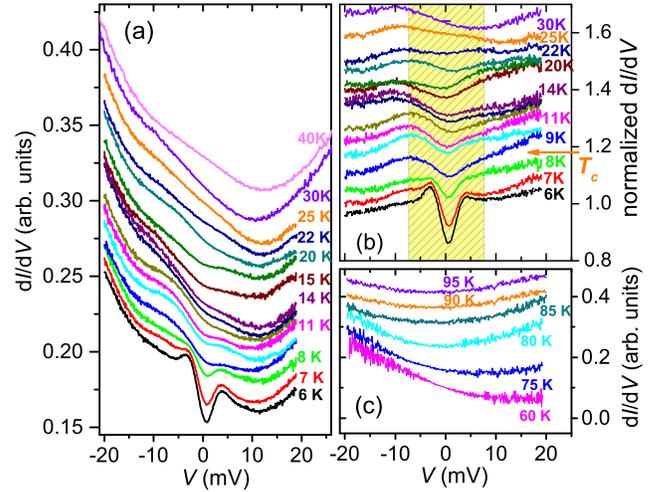}
\caption{(a) Raw tunneling spectra in the temperature range 6-40~K. (b) Normalized tunneling spectra in the temperature range 6-30~K. The yellow shaded region represents the energy scale of $T^{**}$. (c) Raw tunneling spectra in the temperature range 60-95~K. The spectra are shifted vertically for clarity. For all the spectroscopy measurements, $V_{g}$ = 0.02 V  and $I_{sp}$ = 0.6 nA before opening the STM feedback loop. The bias modulation amplitude was set to 0.1~m$V_{rms}$. Reproduced from Ref. \cite{Ros2015}, copyright (2015) by American Physical Society.}
\end{figure}
 
Albeit the above discussed controversial results concerning the origin of the tetragonal-orthorhombic phase transition, the two different Fe-Fe distances ($d_\mathrm{Fe1} \neq d_\mathrm{Fe2}$) in the orthorhombic phase, see Fig. 4, have the following consequence: due to the non-equal Fe-Fe distances, the orbital occupancies $\Braket{n}$ of the $d_{xz}$ and $d_{yz}$ orbitals are not equal, \textit{i.e.}, $\Braket{n_{xz}}\neq  \Braket{n_{yz}}$, usually described as orbital order or nematicity. However, such a preferential partial occupation of one of the $d_{xz}$/$d_{yz}$ orbitals might induce additional reconstructions of the Fermi surface, $e.g.$ a Peierls-type dimerization or an onset of some type of density wave \cite{Ros2015}. Both types would spontaneously break the translational symmetry of the lattice. An early electron diffraction experiment \cite{Mc2009b} on FeSe indeed reported the actual symmetry of FeSe below 20~K is lower than that of $Cmma$. Based on telltale signs in the magnetoresistance and Hall effect measurements (Fig. 5) as well as scanning tunneling spectra (Fig. 6) on the single crystals grown by chemical vapor transport \cite{Koz2014}, we identified the emergence of an incipient ordering mode and its nucleation  \cite{Ros2015}. The onset temperature $T^{*} \approx$ 75~K of this ordering mode is clearly lower than the structural transition temperature $T_{s}$ = 87~K, and therefore must be of different origin. The temperature $T^{*}$ was identified from the temperature at which (i) an onset of spin fluctuations was found in magnetic susceptibility and NMR measurements; (ii) an enhancement of positive magnetoresistance was observed; (iii) a deviation from the Kohler's scaling $\mathcal{F}[H/\rho(0)]$, which could be seen in Figs. 4 (a) and (b); (iv) the Hall resistivity $\rho_{xy}(H)$ deviates from linearity, Fig. 5(c); (v) an inflection in the temperature dependence of the initial Hall coefficient $R_{H\rightarrow0  }(T)=\frac{\delta \rho_{xy}(T)}{\delta H}\rvert_{H\rightarrow0}$ could be obtained, Fig. 5(d); (vi) the tunneling spectra develop an asymmetry, Fig. 6 (c), suggesting a non-compensation of occupied (electron-like) and unoccupied (hole-like) states. Here we would like to mention that several other groups reported a similar behavior of the Hall coefficient $R_{H\rightarrow0  }(T)$ measured on single crystals grown by different methods \cite{Huy2014,Kas2014,Wat2015b,Lei2012,Sun2016} indicating that the behavior is intrinsic, and not sample dependent. While Refs. \cite{Huy2014,Kas2014,Wat2015b} attributed the negative values of $R_{H\rightarrow0  }(T)$ at low temperature to an enhanced mobility of the electrons in a compensated metal, Refs. \cite{Ros2015,Lei2012} considered an increased number of negative charge carriers at low temperatures. The latter interpretation is also supported by a negative sign of the Seebeck coefficient found at low temperatures for FeSe \cite{Mc2009,Kas2014,Son2011}. A screening for further evidences for an emerging ordering mode at $T^{*}$ in other measurements reported in literature revealed that such signs were not identified as a separate feature different from the nematicity. For instance, a negative value of the Seebeck coefficient was observed at $T < T_{s}$, see supplementary information in Ref. \cite{Kas2014}. Further, the sign of the elastoresisitance tensor  $m_{66}$ changes sign at 65~K \cite{Wat2015} and the resistivity anisotropy peaks at around 70~K \cite{Tan2015}. In both Refs. \cite{Wat2015} and \cite{Tan2015}, the behavior was attributed to anisotropic inelastic scattering originating from the enhanced spin fluctuations at the respective temperatures. 

In our measurements \cite{Ros2015}, we speculated about another temperature scale $T^{**} \approx$ 22 - 30~K, marked by the opening up of a partial gap of about 8 meV (Fig. 6 b) in tunneling spectra as well as a recovery of Kohler's scaling (Fig. 5a). Based on these results we suggested that $T^{*}$ represents the onset of an incipient order associated with enhanced spin fluctuations. Static nucleation of this mode below a second temperature $T^{**}$ appears to result in a coupling between electronic charge, orbital, and pocket degrees of freedom \cite{Jian2011} at this temperature which is discernible in anomalies of transport data and tunneling spectra. The temperature $T^{**}$ may also be related to the translational symmetry breaking found in electron diffraction experiments \cite{Mc2009b}.  
 
\subsection{Superconductivity}  

When a new superconducting compound is discovered, typically the two main foci of research are i) how to further enhance the transition
temperature ?, and ii) what is the superconducting pairing mechanism ? In the case of FeSe, $T_{c}$ has been enhanced from 8.5~K to 37~K by the application of pressure \cite{Miz2008,Med2009,Mar2009,Ima2009,Sid2009}, and to 65 - 100 ~K by growing FeSe monolayers on SrTiO$_{3}$ substrates \cite{Yan2012,He2013,Tan2013,Ge2015,Fan2015}. The superconducting pairing mechanism, however, still remains elusive. The symmetry of the superconducting order parameter is still under dispute \cite{Hir2011,Hir2016}, but most
researchers favor an unconventional $s\pm$ symmetry with a sign change of the order parameter
between the hole and the electron bands mediated by either nesting-induced spin \cite{Maz2008,Kur2008,Wan2009} or orbital fluctuations \cite{Kon2010}. All these theories are based on multi-orbital models. For FeSe, the density functional calculations yield two intersecting elliptical electron Fermi surfaces at the corner of the Brillouin zone and two concentric hole cylinders at the zone center \cite{Sub2008}. Nevertheless, experimental observation of the Fermi surfaces of FeSe by ARPES and quantum oscillation experiments deviate significantly from those obtained by band structure calculations. 
Quantum oscillation experiments at low temperatures have detected extremely small Fermi surfaces \cite{Ter2014,Wat2015,Wat2015b,Aud2015}. Several ARPES experiments detected only one small hole pocket at the zone center and one electron pocket at the zone corner \cite{Mal2014,Shi2014,Zha2015,Wat2015,Nak2014,Suz2015}. Concerning the hole pocket, the quantum oscillations are in agreement with the ARPES experiments, but the electron pocket measured in the two experiments appear to be different. Based on their combined results of the quantum oscillations and ARPES, Watson \textit{et~al.} \cite{Wat2015,Wat2015b} suggested a scenario with two electron pockets and one hole pocket crossing the Fermi energy. In any case, experimentally detected Fermi surfaces so far did not provide any evidence for nesting of the Fermi surfaces in FeSe. 

\begin{figure}[t]
\includegraphics[width=7.5 cm,clip]{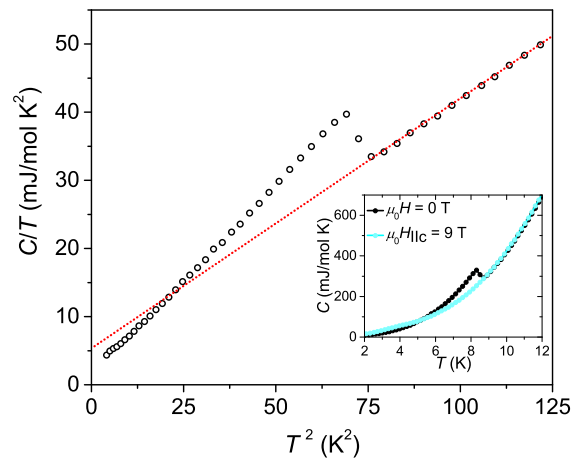}
\caption{Low-temperature specific heat of a $\beta$-FeSe single crystal grown by chemical vapor transport \cite{Koz2014}. The red
dotted line is the extrapolation of the 0 T high temperature data. 
The inset presents $C_p(T)$ measured at 0 and 9 T with magnetic
field parallel to the $c$-axis \cite{Koz2015}.}
\end{figure}

Now we turn to some experiments which shed light on the symmetry of the superconducting order parameter in FeSe. In Fig. 7, the heat capacity measurement $C_p(T)$ on a FeSe single crystal grown by chemical vapor transport \cite{Koz2014} is presented. The inset shows the $C_p(T)$ measured in fields of 0 and 9 T with the field applied parallel to the $c$-axis. By fitting $C_p(T)$ taken at 9 T in the temperature range 5 - 13~K to $\gamma T + \beta T^3$, the coefficient $\gamma$ of the electronic contribution to the specific heat in the normal state is estimated to be 5.11(11) mJ/mole K$^2$. This value is also comparable to the extrapolation of the zero-field, high-temperature ($T>T_{c})~C_p(T)$-data (red dotted line in Fig. 7). By the balance of entropy around the superconducting transition, the dimensionless specific-heat jump at $T_{c}$ is determined as $\mathrm{\Delta} C /\gamma  T_{c} = 2.0(1)$ \cite{Koz2015}. This value is significantly higher than the Bardeen-Cooper-Schrieffer (BCS) value of 1.43 for the weak electron-phonon coupling scenario \cite{Bar1957}. A more detailed analysis of the specific heat data by Lin \textit{et~al.} \cite{Lin2011} found an isotropic gap of $\mathrm{\Delta_{0}}$ = 1.33 meV on the hole Fermi sheets and an extended $s$-wave gap $\mathrm{\Delta =\Delta_{e}}(1+\alpha~\mathrm{cos2}\phi)$ with $\mathrm{\Delta_{e}}$ = 1.13 meV and $\alpha$ = 0.78 on the electron Fermi sheets. Further, the London penetration depth
$\lambda_{ab}(T )$ calculated from the temperature ($T$) dependence of the lower critical fields could not be fitted to a single-gap BCS model \cite{Abd2013}. This $T$-dependence could be described by using either a two-band model with $s$-wave-like gaps of magnitudes $\mathrm{\Delta_1}$ = 0.41$\pm$0.1 meV and  $\mathrm{\Delta_2}$ = 3.33$\pm$0.25 meV or a single anisotropic $s$-wave order parameter \cite{Abd2013}. Two-gap behavior was also reported based on the superfluid density measurements using muon-spin rotation \cite{Kha2010}. Thermal conductivity measurements on single crystalline FeSe also report two node-less superconducting gaps \cite{Dong2009,Hop2016}. In any case, these bulk measurements support two superconducting gaps without nodes. In contrast, surface sensitive scanning tunneling spectroscopic (STS) measurements on FeSe, both in the single crystalline \cite{Kas2014} and the thin film form \cite{Song2011}, detected a ``V''-shape spectra in the superconducting state, which was interpreted as being indicative of the presence of nodes. Interestingly, it was also shown by STS measurements that the tunneling spectra display a full gap at the twin boundaries \cite{Wata2015}. However, our most recent combined STS and specific heat studies conducted on single crystals grown by pure chemical vapor transport \cite{Koz2014} strongly support multigap node-less superconducting gap structure through out the FeSe material \cite{Jiao2016}.  

One recent experiment that connected both nematicity and superconductivity to spin-fluctuations is inelastic neutron scattering on FeSe single crystals \cite{Wan2016}. They found a spin resonance of 4 meV in the superconducting state. The temperature dependence of this mode behaved like an order parameter, thus the authors argued that the spin-resonance is coupled to the onset of superconductivity. The resonance energy was compared with the electron-boson coupling mode at similar energies reported based on STS experiments \cite{Song2014}. Based on these arguments, these authors suggested a spin-fluctuation-mediated, sign-changing pairing mechanism for FeSe. Further, antiferromagnetism coexisiting with superconductivity has been reported in FeSe upon application of pressure \cite{Ben2010,Ben2012,Ter2015,Ter2016}. 

From the above overview of the current experimental status it can be discerned that unequivocal conclusions on the origin of nematicity, the pairing mechanism as well as the order parameter symmetry of FeSe can still not be drawn. More experiments on high-quality single crystals are required for testing scenarios such as orbital or spin-fluctuation driven nematicity and/or superconductivity. It is also worthwhile to consider phonons as the driving force of the structural transition, and then investigate how this broken $C_4$ symmetry affects the underlying electronic system.

\section{F\lowercase{e}$_{1+y}$T\lowercase{e}}

The compound Fe$_{1+y}$Te has the same tetragonal structure at room temperature as FeSe, but forms only in the presence of excess Fe ($y$). Fe$_{1+y}$Te has also remarkably different physical properties. It does not exhibit superconductivity in the bulk form. Instead, it shows a very rich interplay of localized and itinerant magnetism \cite{Zal2011,Sto2014,Duc2014}. Moreover, magnetostructural  phase transitions have been observed when tuning parameters such as temperature, composition, or pressure were varied \cite{Bao2009,Li2009,Rod2011,Ros2011,Zal2012,Koz2012,Koz2013,Rod2013,Koz2016}. As a digression, while $T_{c}$ can be dramatically increased in FeSe monolayers \cite{Yan2012,He2013,Tan2013,Ge2015,Fan2015}, superconductivity has been found in Fe$_{1+y}$Te when fabricated in the form of thin films \cite{Han2010} or heterostructures \cite{He2014}.

\subsection{Synthesis and characterization}

\begin{figure}[t]
\includegraphics[width=7.5 cm,clip]{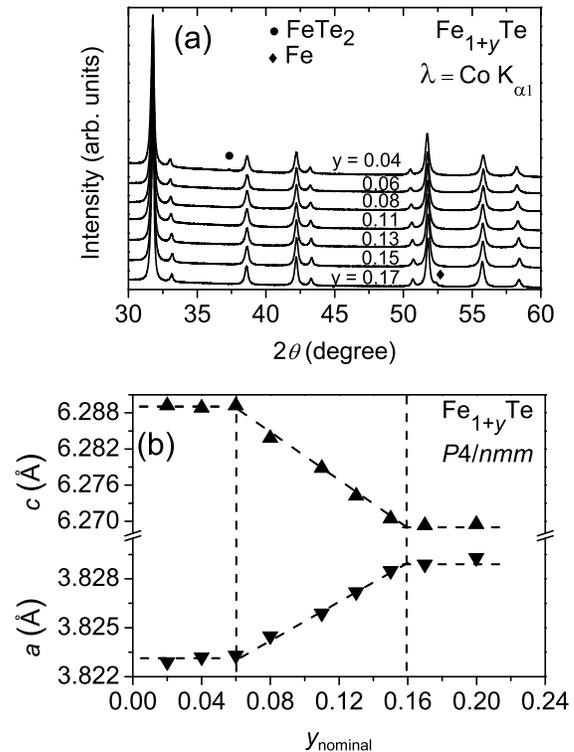}
\caption{X-ray diffraction diagram of Fe$_{1+y}$Te with nominal composition for $y = 0.04-0.17$. The main phase is tetragonal at room temperature. (Impurity phases; FeTe$_{2}$ marked by \textbullet~and elemental Fe by $\blacklozenge$) (b) Lattice parameters at room temperature in dependence on the nominal composition $y$. The error bars here are smaller than the size of the symbols. Reproduced from Ref. \cite{Koz2013}, copyright (2013) by American Physical Society.}
\end{figure}
Polycrystalline Fe$_{1+y}$Te samples are typically synthesized utilizing a solid state reaction of
Fe and Te pieces with different amounts of iron contents in the range 0.02 $\leq y \leq$ 0.20.
Mixtures of the target composition were placed in a glassy carbon crucibles with lid
and sealed in quartz ampoules under vacuum. Starting materials were first heated up
to 700$^\circ$C at a rate of 100$^\circ$C/h and kept there for one day. Subsequently, the temperature was
slowly increased to 920$^\circ$C in order to complete the reaction between Fe and Te and to
obtain homogeneous products. After maintaining a temperature of 920$^\circ$C for 48 h, samples were cooled down
to 700$^\circ$C and annealed for 24 h in order to avoid the formation of a high-temperature
modification of Fe$_{1+y}$Te which was reported above 800$^\circ$C \cite{Ros1974,Isp1974,Isp1974b}.
Initially, the chemical composition of  Fe$_{1+y}$Te was reported to be in the range 
0 $\leq y \leq$ 0.3 \cite{Chi1955,Oka1990,Miz2012}. A more detailed investigation by Koz \textit{et~al.}
found that the actual phase stability range for tetragonal Fe$_{1+y}$Te is only for compositions
with 0.060(5) $< y <$ 0.155(5) \cite{Koz2015,Koz2013}. In Fig. 8(a), the 
powder x-ray diffraction (PXRD) patterns of Fe$_{1+y}$Te (y = 0.04, 0.06, 0.08, 0.11, 0.13, 0.15, and 0.17
are presented. The tetragonal Fe$_{1+y}$Te is the main phase at room temperature for all studied nominal compositions, and the
main reflections can be indexed on the basis of a tetragonal cell of space group $P4/nmm$. For $y <
0.06$ and $y > 0.15$, FeTe$_{2}$ \cite{Gro1954} and Fe impurities were observed in the PXRD patterns, respectively.
The lattice parameters of tetragonal Fe$_{1+y}$Te at room temperature are presented in Fig. 8(b).
In the composition range  0.060(5) $\leq y \leq$ 0.155(5), the $a$ parameter increases while the $c$ parameter decreases upon increasing Fe
content. The volume of the unit cell does not change significantly with composition, however, the
$c/a$ ratio decreases from 1.645 to 1.638 with increasing $y$ \cite{Koz2015}.

In the following, we discuss the site occupancy of excess Fe.  Gr{\o}nvold \textit{et~al.}
 \cite{Gro1954} first suggested that the excess iron atoms in the Fe$_{1+y}$Te phase are located in
partially occupied interstitial sites. The crystal structure was considered to be an
intermediate type between tetragonal PbO (the B10 type) and Fe$_{2}$As (the C38 type).
The arrangement in the B10 structure as described for $\beta$-FeSe is a quadratic net of
iron atoms, which together with the tellurium atoms form square pyramids with sharing
edges, and with apices alternately above and below the net of iron atoms. The excess Fe atoms (Fe2) occupy an additional site in the tellurium plane and
convert a pyramid into an octahedron without appreciably disturbing the original
grouping. In space group $P4/nmm$, the Fe1 and Te atoms fully occupy the $2a$ and
$2c$ sites, respectively, while the Fe2 atoms randomly occupy a second $2c$ site. These
conclusions have been confirmed using x-ray and neutron diffraction measurements \cite{Bao2009,Li2009,Fru1975,Sal2009,Liu2009}.

\subsection{Single crystal growth}

The single crystals of Fe$_{1+y}$Te were generally grown either by a self flux method \cite{Rod2011,Miz2012b}, by the horizontal or vertical Bridgman method \cite{Sal2009,Wen2011,Che2014}, or by chemical vapor transport \cite{Koz2015}. The growth started either from the mixture of elemental Fe and Te \cite{Miz2012,Che2014} or by taking premade powder of Fe$_{1+y}$Te \cite{Rod2011,Koz2015}. Because of the non-stoichiometry, it proved to be difficult to control the exact Fe content in the single crystals. In the case of the horizontal Brigdman method \cite{Che2014,Che2015}, when the starting composition of Fe:Te was kept at 1.04:1, crystals of Fe$_{1.06}$Te could be grown \cite{Che2015}. For the growth of crystals with higher Fe content, the starting ratio of Fe:Te was kept at 1.09:1. Note that both Fe$_{1.11}$Te and Fe$_{1.12}$Te crystals were extracted from a single growth, but from different parts of the ampoule. The exact compositions were determined only after a detailed chemical and physical analysis. The chemical vapor transport also had very similar problems in controlling the stoichiometry. Different temperature gradients between 700 and 800$^\circ$C, and different transport additives, such as NH$_{4}$I, GaI$_{3}$, and TeCl$_{4}$, did not help to optimize the composition of the end products. However, a trend was observed during the experiments with I$_{2}$. The single crystals contain roughly ~2 \% less iron than the starting composition. For example, polycrystalline materials with nominal composition  Fe$_{1.15}$Te yielded single crystals with composition  Fe$_{1.13}$Te. Good quality single crystals of  Fe$_{1+y}$Te ($0.11 \leq y \leq 0.14$) could be easily grown under respective conditions. However, when the starting composition was y = 0.11, the grown crystals were found to be a mixture of the tetragonal phase with different compositions.  With further decrease in the starting composition, $i.e.$ y $<$ 0.11, the grown crystals formed in another phase with chemical composition of Fe$_{1-x}$Te ($0.08 \leq x \leq 0.15$). This new phase could not be identified from the PXRD pattern. However, the compositions estimated from energy dispersive x-ray (EDX) measurements were in good agreement with the $\delta$-phase which was reported between 636 and 809$^\circ$C in the Fe-Te phase diagram \cite{Oka1990}. Hence, in chemical vapor transport with I$_{2}$ as a transport additive, only crystals of the compositions Fe$_{1.11}$Te, Fe$_{1.12}$Te and Fe$_{1.14}$Te could be grown. In Fig. 9, images of such crystals are presented. Alternatively, Rodriguez \textit{et~al.} reported that I$_{2}$ can be used as an oxidant in the topotactic deintercalation of interstitial iron in  Fe$_{1+y}$Te and Fe$_{1+y}$Te$_{0.7}$Se$_{0.3}$ \cite{Rod2010,Rod2011b}. According to their results, iodine reacts with iron at 300$^\circ$C to form FeI$_{2}$. This might explain the composition difference of starting and end products in our chemical vapor transport reactions. Using the deintercalation procedure, crystals of Fe$_{1.051(5)}$Te could be obtained from Fe$_{1.118(5)}$Te \cite{Rod2010}.

\begin{figure}[t]
\includegraphics[width=7.5 cm,clip]{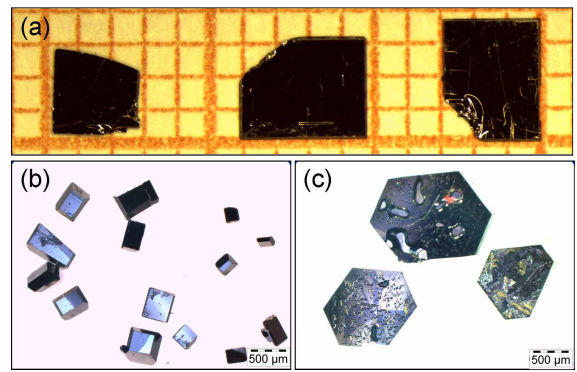}
\caption{Images of single crystals grown by chemical vapor transport \cite{Koz2015}. (a) Big plate-like
Fe$_{1+y}$Te single crystals, (b) typical tetragonal Fe$_{1+y}$Te single crystals grown in a batch, (c)
crystals with hexagonal shape likely belong to the $\gamma$-phase with a composition FeTe$_{1.2}$.}
\end{figure}

\subsection{Structural and magnetic properties}
Based on neutron scattering experiments on Fe$_{1+y}$Te, Bao \textit{it~al.} \cite{Bao2009} made two important observations. (i) The magnetic and structural properties of Fe$_{1+y}$Te are extremely sensitive to the amount of excess Fe ($y$). For example, Fe$_{1.076}$Te undergoes a structural phase transition to a monoclinic ($P2_1/m$) bicollinear antiferromagnetic phase whereas for Fe$_{1.141}$Te, a structural phase transition to an orthorhombic ($Pmmn$) phase with an incommesurate antiferromagnetic structure was observed.
(ii) The bicollinear antiferromagnetic structure of Fe$_{1.076}$Te has a wave vector along $\mathrm{{\bf{q_{AFM}}}}=[\frac{1}{2},0,\frac{1}{2}]$. When defined in the 1-Fe Brillouin zone, this corresponds to ($\pi$/2,$\pi$/2) ordering.  Along this wave vector, no nesting has been observed either in the ARPES experiments \cite{Xia2009,Lin2013} or in the density functional calculations \cite{Sub2008}. The results of neutron scattering indicated the presence of local moments. Indeed, above the antiferromagnetic ordering temperature $T_{N}$, the magnetic susceptibility $\chi(T)$ follows a Curie-Weiss behavior \cite{Koz2016}. The value of the total spin $S$ calculated from the Curie constant was found to be $S$ = 3/2, which is also consistent with the value obtained from the inelastic neutron scattering \cite{Zal2011,Sto2014}. Further, below $T_{N}$, the ordered magnetic moment corresponds to $S$ = 1 \cite{Zal2011,Bao2009}. The values of $S$ both above and below $T_{N}$ are clearly lower than $S$ = 2 expected for Fe$^{2+}$ in the tetrahedral coordination. The electrical properties of Fe$_{1+y}$Te shows nonmetallic character in resistivity for temperatures above $T_{N}$ with $d\rho/dT < 0$, indicative of charge carrier incoherence near the Fermi level \cite{Liu2009,Chen2009}. ARPES measurements on Fe$_{1.02}$Te revealed a sharp feature near the Fermi energy $E_{\mathrm{F}}$ below $T_{N}$, suggesting the appearance of coherent charge carriers in the AFM phase \cite{Liu2013b}. A similar behavior was also observed in scanning tunneling spectroscopy \cite{Ros2012}. These results suggest an entanglement of localized and itinerant electrons in Fe$_{1+y}$Te \cite{Zal2011,Tur2009}.

\begin{figure}[t]
\includegraphics[width=8.3 cm,clip]{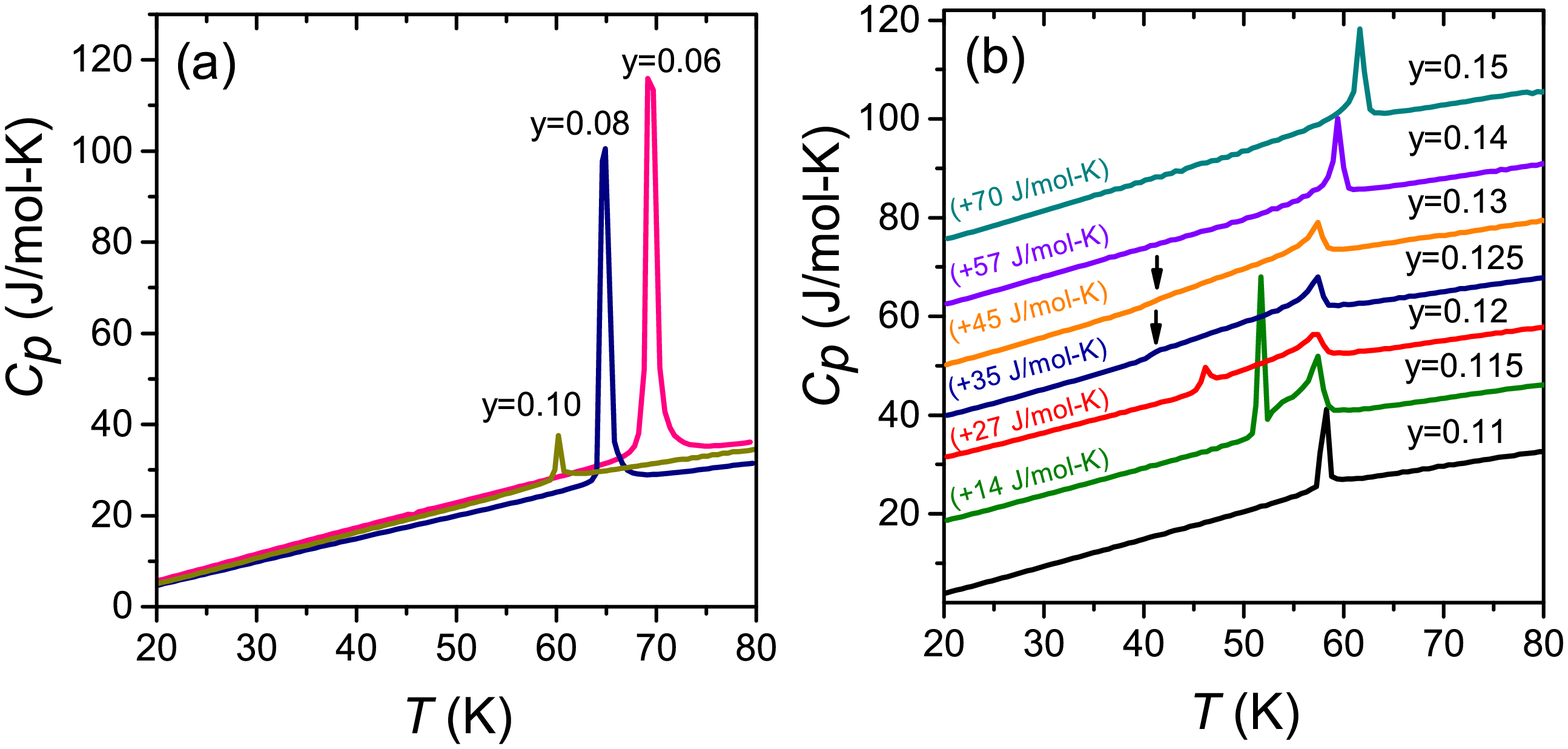}
\caption{Specific heat $C_p(T)$ of Fe$_{1+y}$Te for (a) $y$ = 0.06, 0.08, and 0.10 and (b) $y$ =
0.11-0.15. The latter are shifted for clarity. Arrows show the
disappearing first-order phase transition upon increasing Fe composition \cite{Koz2015}.}
\end{figure}

\begin{figure}[t]%
\includegraphics*[width=8.5 cm, clip]{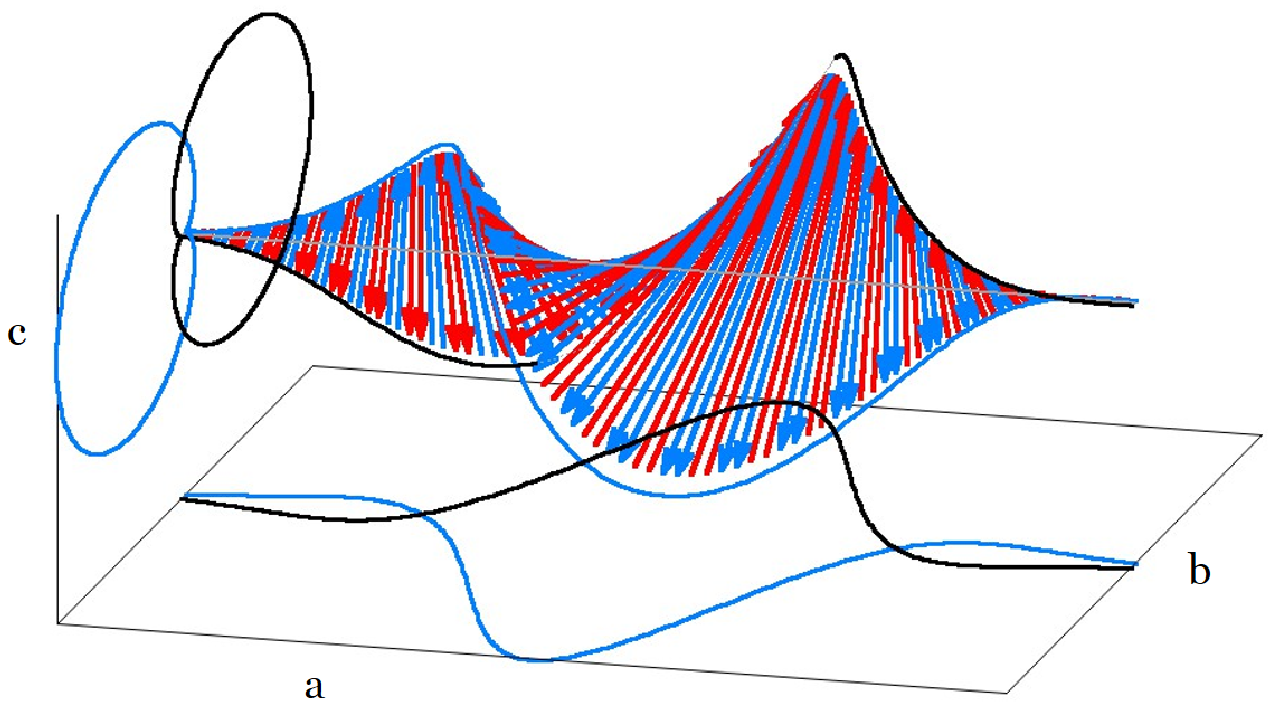}
\includegraphics*[width=8.5 cm, clip]{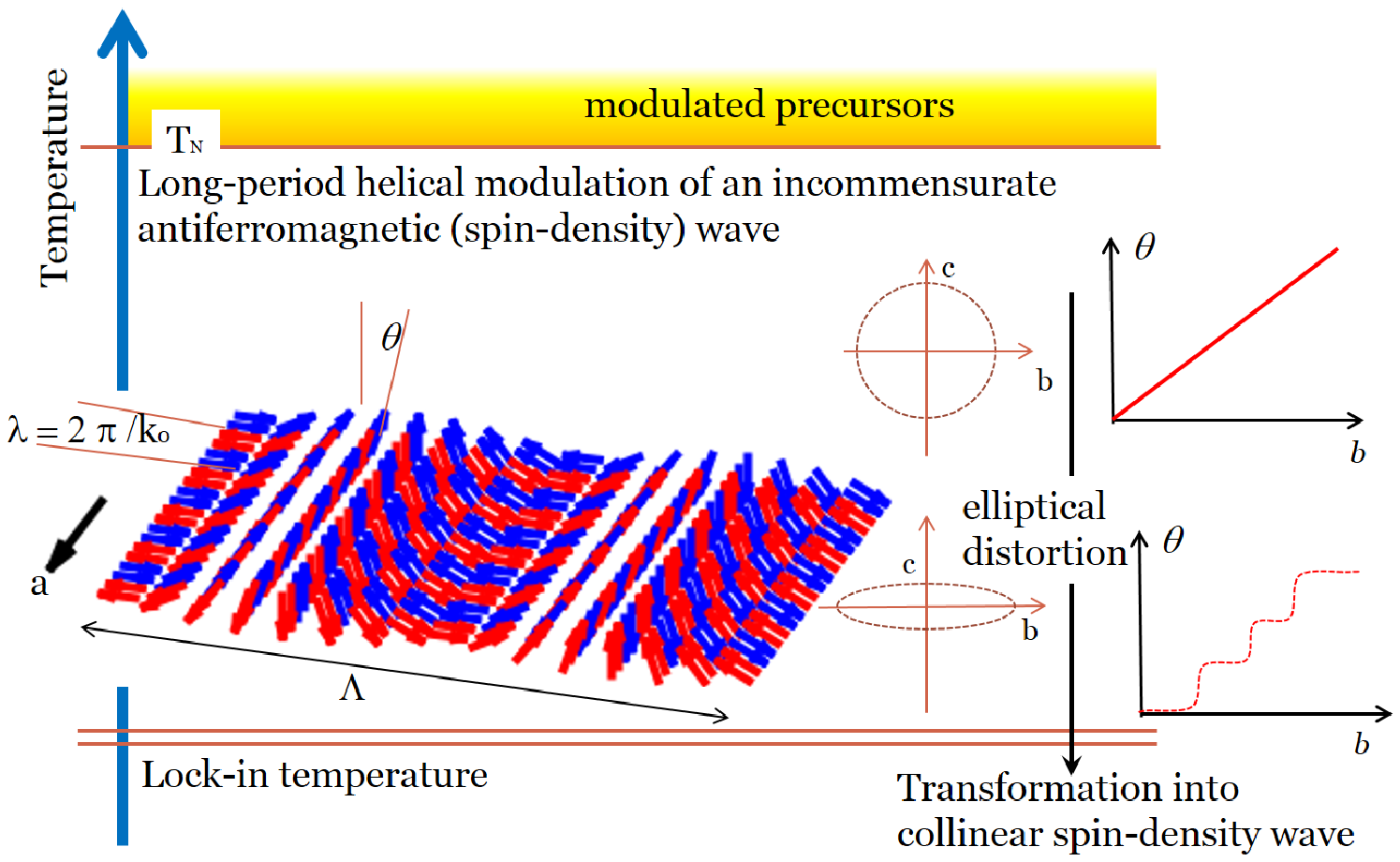}
\caption{Top panel: picture of a possible solitonic precursor fluctuation of an incommensurate spin-density wave-like order above the ordering temperature.
Direction and amplitude of the approximate two-up-two-down order are twisted and modulated, respectively,  over long lengths (here in direction of the propagation vector of the spin density wave parallel to $a$).
Bottom panel: pictorial representation of the magnetic phase diagram in Fe$_{1+y}$Te in the range $0.11 \le y \le 0.13$.
Above $T_N$ a dense solitonic liquid state precedes the long-rang order.
At $T_N$ the incommensurate long-range order is established, that can be twisted transverse to the propagation direction $a$ over a long period $\Lambda$ in the $bc$-plane, here taken in $b$-direction. The  rotation angle $\theta$ vs $b$ parametrizes a cycloidal twisting of the spin-density wave.
Magnetic anisotropies cause elliptic distortions of the harmonic twisting  and the formation of a regular soliton lattice in the  magnetically ordered  state upon lowering temperature, which finally disappears at the lock-in temperature where a collinear spin-density wave state forms the magnetic ground-state. Figure courtesy by U. K. R\"o{\ss}ler.
}
\label{Fig10}
\end{figure}

More detailed low temperature neutron as well as SXRD experiments revealed three different types of magnetic and structural behavior in Fe$_{1+y}$Te  depending on the amount of excess Fe \cite{Rod2011,Ros2011,Zal2012,Koz2013,Rod2013,Miz2012,Che2014b}. In the compositional range $y < 0.11$, Fe$_{1+y}$Te undergoes a simultaneous first-order phase transition to a monoclinic ($P2_1/m$) as well as bicollinear antiferromagnetic phase. For the intermediate composition $0.11 \leq y \leq 0.13$, two coupled magneto-structral phase transitions were observed \cite{Ros2011,Zal2012,Koz2013,Miz2012,Che2014b}. Upon cooling, the system undergoes a continuous phase transition to an orthorhombic ($Pmmn$) incommesurate antiferromagnetic phase, followed by a first-order lock-in transition to the monoclinic bicollinear antiferromagnetic phase. It was, however, found that the low-temperature phase is a mixture of both, the orthorhombic and the monoclinic phases \cite{Koz2013,Rod2013}. For $y > 0.13$, the material undergoes a single phase transition to the orthorhombic incommensurate antiferromagnetic phase \cite{Rod2011,Sto2011}. This behavior is nicely corroborated by measurements of the specific heat $C_p(T)$ of Fe$_{1+y}$Te, Fig. 10. The sharp peaks found in $C_p(T)$ for $y < 0.11$ confirm a first-order phase transition. For $0.11 \leq y \leq 0.13$, two peaks can be discerned in $C_p(T)$. The transition at higher temperature has a $\lambda$-like shape in $C_p(T)$ suggesting a continuous phase transition, while the peaked shape of the low-temperature transition is first-oder like. For $y > 0.13$, once again a single $\lambda$-like peak is found in $C_p(T)$ suggesting a continuous nature of the phase transition.   

The bicollinear antiferromagnetic ordering can be captured by a frustrated $J_{1}$-$J_{2}$-$J_{3}$ Heisenberg model \cite{Ma2009}. However, the antiferromagnetic ordering for $y\geq$ 0.11 is rather unconventional. For this compositional range, a random field Potts model \cite{Zal2012} or models involving several propagation vectors (plaquette ordering pattern) were proposed \cite{Duc2012,Ena2014}. Based on a phenomenological Landau theory, U. K. R\"o{\ss}ler \cite{Mat2015} predicted the presence of liquid-like mesophases as precursors in Fe$_{1+y}$Te. These precursors are composed of solitonic amplitude-modulated states, Fig. 11 (top panel). A dense amorphous condensate of such particle-like states is formed as a precursor at higher temperatures $T > T_{N}$ before a coherent magnetic long-range ordering takes place. Evidence for such magnetic precursors have been found in inelastic neutron scattering \cite{Sto2011,Par2012} as well as M\"ossbauer spectrosocpy \cite{Mat2015}. At the ordering temperature $T_{N}$, a long-period modulation of a helical antiferromagnetic spin density wave is established. Upon decreasing temperature, the modulation direction is determined  by a strong anisotropy, thereby making the helix more and more elliptical, until a bicollinear antiferromagnetic order is achieved at the lock-in transition. This process is pictorially represented in Fig. 11 (lower panel). The transition from a helical incommensurate spin density wave to bicollinear antiferromagnetic order was also observed in neutron diffraction experiments \cite{Bao2009,Rod2011,Zal2012,Fob2014}.   

\subsection{Effects of pressure}

\begin{figure}[t]
\includegraphics[width=8.5 cm,clip]{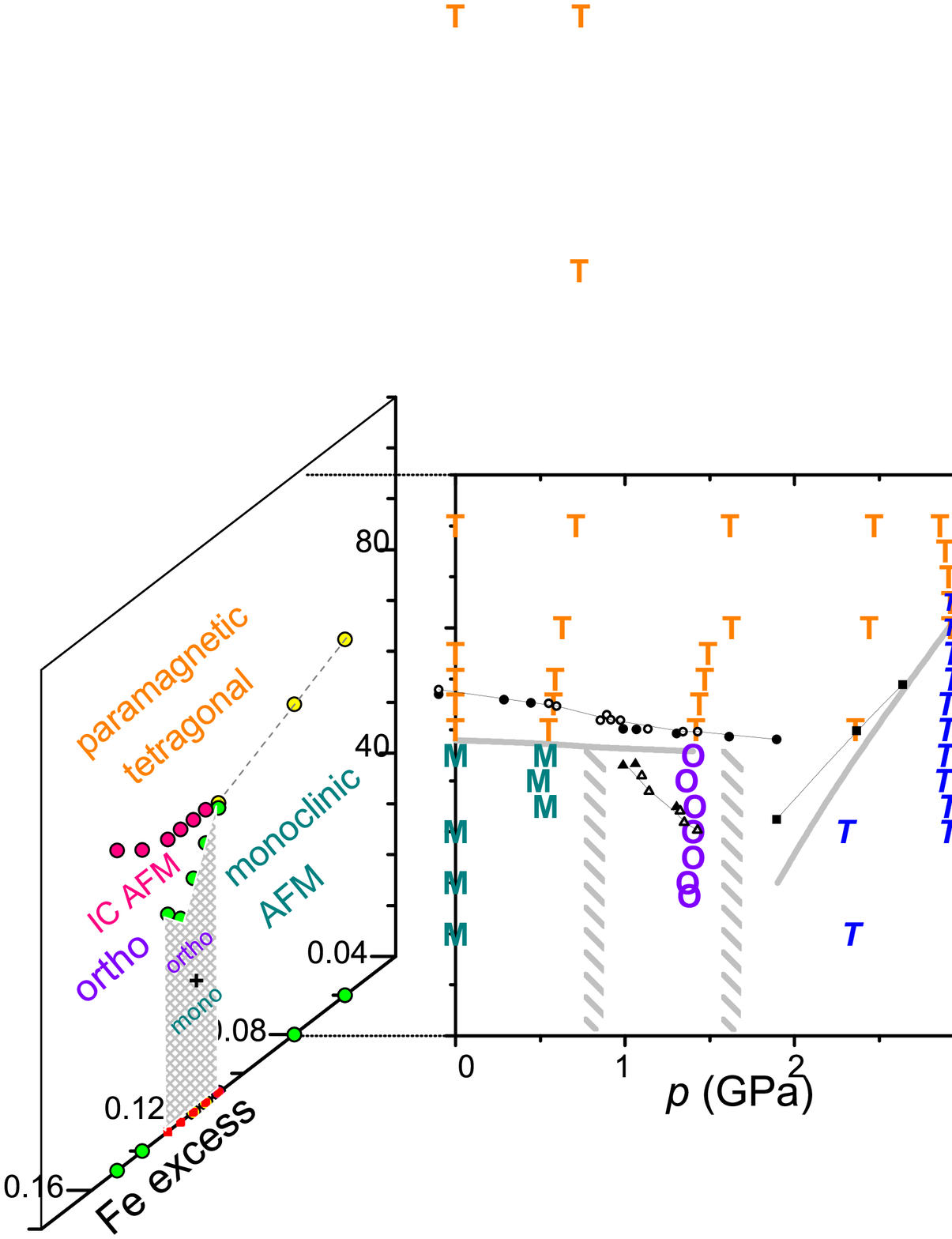}
\caption{Temperature-pressure-composition phase diagram for the Fe$_{1+y}$Te system. Symbols T, O and M mark temperatures and pressures of our XRD measurements revealing tetragonal, orthorhombic and monoclinic phases, respectively. The black data points indicate anomalies in resistivity, taken from \cite{Oka2009} for samples Fe$_{1.086}$Te. AFM and IC AFM stand for antiferromagnetic
and incommensurate antiferromagnetic phase, respectively.}
\end{figure}

The temperature-pressure phase diagram of Fe$_{1+y}$Te is remarkably similar to the temperature-composition (excess Fe) phase diagram as can be seen from Fig. 12. Based on resistivity and magnetization measurements, Okada \textit{et~al.} \cite{Oka2009} first identified two pressure-induced phases at low temperatures in FeTe$_{0.92}$ (Fe$_{1.086}$Te). In order to identify the phases, Koz \textit{et~al.} \cite{Koz2012} performed high-pressure SXRD on Fe$_{1.08}$Te.  At ambient pressure, Fe$_{1.08}$Te undergoes simultaneous first-order structural and magnetic phase transitions, $i.e.$ from the paramagnetic tetragonal ($P4/nmm$) to the antiferromagnetic
monoclinic ($P2_1/m$) phase. At a pressure of 1.33 GPa, the low-temperature structure adopts
an orthorhombic symmetry in the space group $Pmmn$. More importantly, for pressures of 2.29 GPa and higher, a symmetry-conserving
tetragonal-tetragonal phase transition has been identified from a change in the $c/a$ ratio of the lattice parameters.
From the high-pressure magnetization measurements, the high-pressure, low-temperature tetragonal phase was found to be ferromagnetic \cite{Ben2013}. Interestingly, unlike the parent compounds of Fe-pnictides, no superconductivity was observed in Fe$_{1+y}$Te up to a pressure of 19 GPa \cite{Oka2009}. The close resemblance of the temperature-composition and the temperature-pressure phase diagrams suggests a strong magneto-elastic coupling between the magnetic and structural order parameters in Fe$_{1+y}$Te.

\section{Concluding remarks}
In this article, we have reviewed the synthesis and properties of two isostructural materials belonging to the family of Fe-chalcogenides. While FeSe is an itinerant non-magnetic compound, 
Fe$_{1+y}$Te displays an interplay of localized and itinerant properties. The magnetic ordering in Fe$_{1+y}$Te is bicollinear antiferromagnetic with the magnetic vector along ($\pi$/2, $\pi$/2) defined in the 1-Fe Brillouin zone. Upon substituting Se for Te, the long-range ($\pi$/2, $\pi$/2) order is suppressed and superconductivity emerges. Interestingly, in Fe$_{1+y}$Te$_{1-x}$Se$_{x}$, a resonance of a soft magnetic mode appears at the wave vector ($\pi$, 0) (\cite{Liu2010}, note that this vector is defined in the 2-Fe unit cell in the original article) and becomes dominant for $x \geq 0.29$. These results support nesting-based theories of superconductivity, which require magnetic fluctuations along the wave vector ($\pi$, 0) which connects hole and electron parts of the Fermi surface. However, the discovery of high-temperature superconductivity in FeSe monolayers \cite{Yan2012,He2013,Tan2013,Ge2015,Fan2015} as well as in Li$_{1-x}$Fe$_{x}$OHFeSe single crystals \cite{Du2016,Zha2016}, in which the hole Fermi surface is found to be absent at the center of the Brillouin zone, pose a serious challenge to the Fermi surface nesting-based theories of superonductivity in Fe-SC.

Alternative theories consider many-body effects and electron correlations. Although the Mott transition is absent in Fe-SC, these materials are depicted as systems with intermediate
correlations \cite{Qaz2009,Si2008,Si2016}. Dynamical mean-field theory (DMFT) calculations provide evidence for such correlation effects in FeSe \cite{Aic2010}. Considering an interplay of electron kinetic energy, Coulomb potential $U$, and Hund's coupling $J_{H}$, these theories find different effective masses for electrons in different $d$-orbitals. 
The ARPES and quantum oscillation experiments on FeSe detect different band renomalization for different $d$-bands \cite{Mal2014,Ter2014,Wat2015,Wat2015b}, in agreement with the DMFT results. Further, according to the DMFT studies, Fe$_{1+y}$Te is considered as the most strongly correlated among all Fe-based superconductors\cite{Yin2011}. Thus, orbital selectivity seems to be relevant in the case of Fe-chalcogenides \cite{Si2016}.

Another issue that has not been theoretically considered as of primary importance for superconductivity in Fe-SC is the role of spin-orbit coupling. Spin-orbit coupling provides a mechanism for the spins to couple to the lattice, thereby giving rise to a large magneto-elastic effect. A recent ARPES study \cite{Bor2015} on several Fe-SCs including FeSe, detected much larger band splitting due to spin-orbit effects than the possible nematic effects. The size of the spin-orbit coupling was found to be of the same order as the superconducting gap in these materials. In the case of Fe$_{1+y}$Te, strong magneto-elastic effects have been observed in magnetostriction experiments \cite{Ros2012}. Further, scanning tunneling microscopy on Fe$_{1+y}$Te detected a one-dimensional stripe structure \cite{Mac2012,Sug2013}, which is at the same wave vector as the magnetic ordering, indicating the presence of a spin-orbit coupling. These results suggest that the couplings of the electronic subsystem to the phonons are equally important as spin and orbital fluctuations. 

Thus, in spite of nearly one decade of intense research on Fe-chalcogenides, the exact nature of the superconducting pairing mechanism still remains an open question. The current availability of large, good-quality single crystals of these materials provide an opportunity to resolve some of the controversies in the near future.

\section{Acknowledgments}
 Financial support from the Deutsche Forschungsgemeinschaft within the priority program SPP1458 is gratefully acknowledged.  We thank U. K. R\"o{\ss}ler for providing Fig. 11.
 We are grateful to Yuri Grin and Liu Hao Tjeng for support during the entire project. Chien-Lung Huang and Lin Jiao are acknowledged for commenting on the manuscript.

%
%

\end{document}